# Foreign Capital and Economic Growth: Evidence from Bangladesh

Ummya Salma[1]*
Md. Fazlul Huq Khan[2]
Md. Masum Billah[3]

*E-mail for correspondence: ummya.salma@bup.edu.bd

**Abstract**

This study aims to examine the relationship between FDI, personal remittance received, and official development assistance and aid in the economic growth of Bangladesh. This study has used time series data on Bangladesh from 1976 to 2021. This study contributes to the existing literature by developing the foreign capital depthless index (FCDI) and examining its impact on Bangladesh's economic growth. The results of the Vector Error Correction Model (VECM) suggest that the economic growth of Bangladesh depends on FDI, remittance, and aid in the long run. However, these variables have no causal relation with GDP in the short run. The relationship between FCDI and economic growth is also positive in the long run. However, the presence of these three variables has a more significant impact on the economic growth of Bangladesh.

Keywords

Foreign direct investment, remittance, aid, economic growth, gross domestic product, vector error correction model, Bangladesh

**Introduction:**

This research aims to investigate how foreign direct investment (hereafter FDI), the inflow of personal remittance, and official development assistance affect the economic growth of Bangladesh. Bangladesh has achieved higher economic growth in recent years. The gross domestic product (hereafter GDP), exports, per capita income, and other indicators show that Bangladesh is one of the fastest-growing developing countries around the globe. The country was the least developed country in its liberation and achieved very little growth in the next two decades. The country was dependent on agriculture, but in the 1990s, the country was shifted to an industrial and service economy. According to the World Bank, Bangladesh is now a developing country with a per capita GDP of $2458 in 2021.

Foreign capital is one of the critical factors in the economic growth of both developed and developing countries. FDI is attracted by both developing and developed countries. Industrially advanced countries attract FDI to increase the speed of industrialization, maintain sustainable economic growth and reduce unemployment (Hussain and Haque, 2016). Moreover, the efficiency of domestic investment is an essential determinant of the effectiveness of FDI in the host country.

Foreign capital is a source of economic development, job creation and transformation for many emerging and developing countries. Foreign capital consists of Foreign Direct Investment (FDI), Personal Remittance and Official Development Assistance (ODA) and Aid. FDI causes the transfer of technology, capital formation assistance, fostering international trade amalgamation, development of a competitive business environment and development of enterprises (Bari, 2013). FDI positively impacts on the host country's economy through the transfer of technology.

[1] Associate Professor, Bangladesh University of Professionals (BUP), Bangladesh
[2] University of International Business and Economics (UIBE), Beijing, China
[3] Joint Director, Bangladesh Bank (The Central Bank of Bangladesh), Bangladesh

Bangladesh has attained above 5 percent GDP growth rate in the last two decades due to the development of export particularly export of readymade garments. According to World Bank data, the country recorded 7.88 percent GDP growth in 2018.

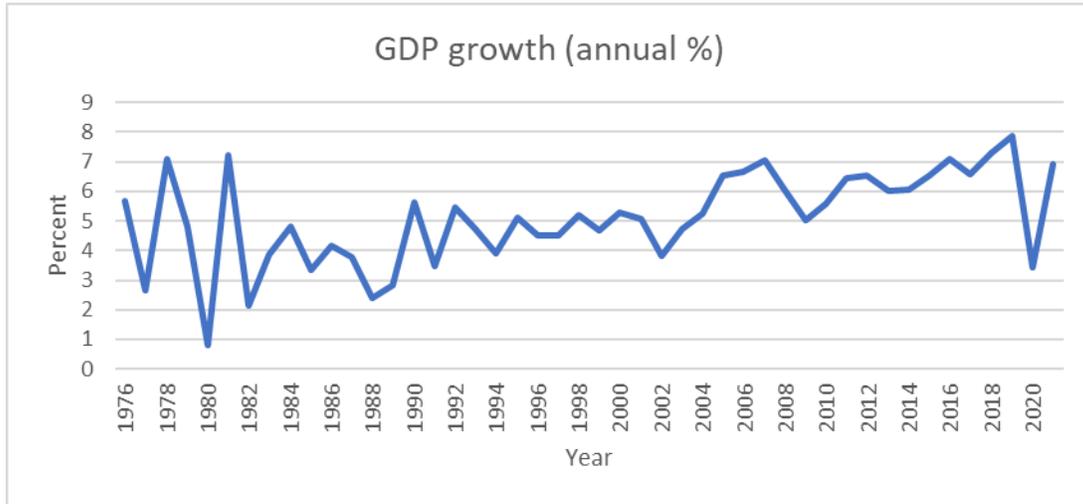

Figure-1: Annual Percentage of GDP growth in Bangladesh

Foreign Direct Investment (FDI) is considered one of country's most significant capital formation components. FDI is a crucial component of capital formation for a developing country like Bangladesh, where public and private savings are scarce, and this investment is highly essential for industrialization in a country (Mujeri and Chowdhury, 2013). According to the OECD, Foreign Direct Investment (FDI) is a cross-border investment where the investor is a resident of one country, invests and establishes an enterprise in another country where the investor has a lasting interest and significant influence over that enterprise. FDI has received extensive attention from policy-makers, researchers, and other development agencies because it contributes to the developing poor people in developing countries. FDI boosts economic growth and encourages more inflow of FDI (Williams, 2017). Developing countries with less savings can accelerate economic growth with the help of foreign capital. Capital inflows can stimulate economic growth by reducing the hindrance the host country faces with its small national savings (Bosworth, 2005). However, short-term capital inflows do not create sustainable growth in developing countries (Williams, 2017). The economic crisis in Mexico in 1994 and the East Asian Economic and financial crisis in the 1990s are attributed to short-term capital inflows in these countries. Bangladesh received only 5.42 million USD as FDI in 1976, but the amount has risen to 1525.31 million USD in 2021. (Anjom, & Faruq, 2023).

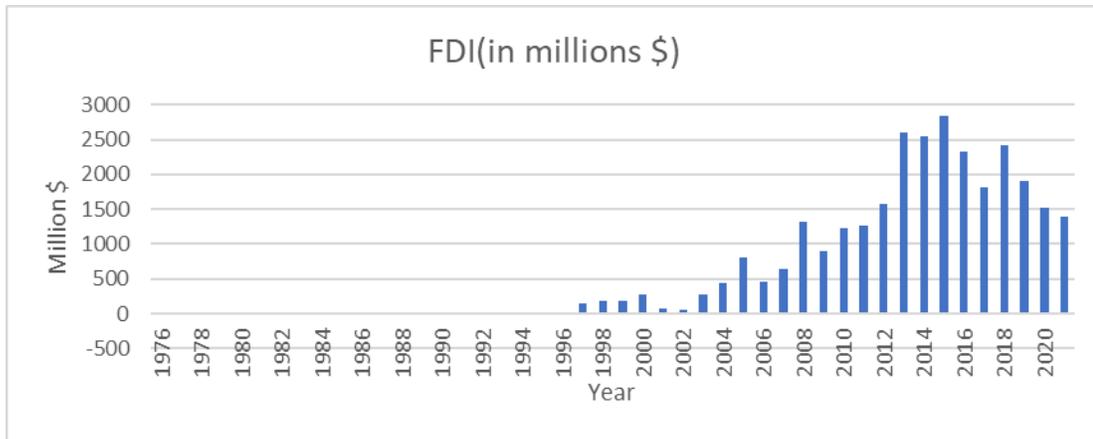

Figure-2: FDI Flows to Bangladesh Over 1976-2021

Remittance plays a significant role in the economic development of developing countries like Bangladesh, where export earnings are less than import-related expenses. According to the International Monetary Fund (IMF), Workers' remittance is the transfer of money by workers living abroad for more than one year to the home country and is recorded in different sections of the balance of payments. According to a study by the World Bank (2018), remittances to developing countries reached a record high of $466 billion in 2017, and were found to have a positive impact on economic growth, poverty reduction, and social development. IMF (2019) found that remittances can help reduce poverty, improve education and health outcomes, and contribute to economic growth in developing countries.

Remittance is one of the important sources of capital and external funding in developing countries (Al-Assaf and Al-Malki, 2014). Personal remittance was 18.76 million USD in 1976. However, in 2021, it was 22202.92 million USD, indicating that a large portion of capital formation depends on remittance.

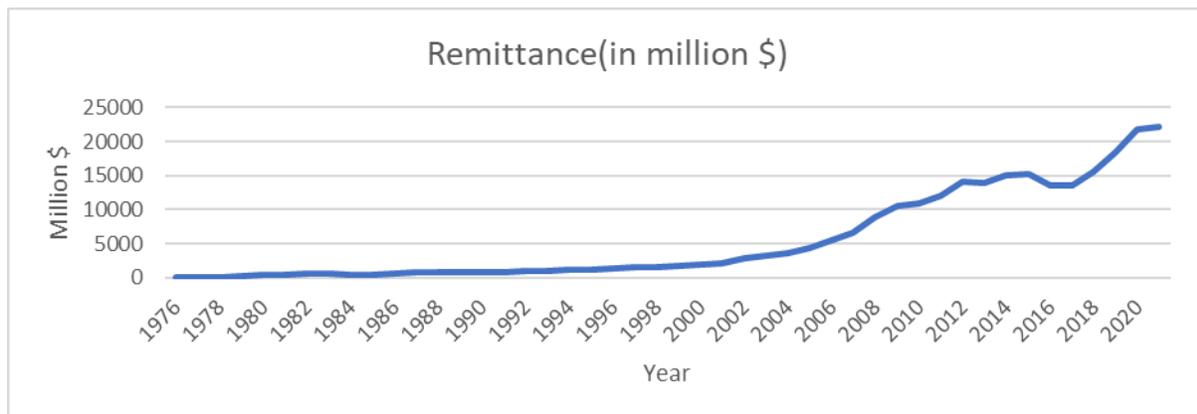

Figure-3: Personal Remittance Flows to Bangladesh Over 1976-2021

Bangladesh has received a large amount of foreign aid and development assistance from different donor agencies and different donor countries since the country is considered one of the poorest countries in the world. The country has a huge population (110 million in 1989) compared to its geographical area (55000 square miles). The per capita income of Bangladesh was 132.46 US Dollars in 1976. The total foreign aid received by Bangladesh in 1976 was 1732.25 million USD, but it was 5041.02 million USD in 2021. After the independence of Bangladesh in 1971, the country received a substantial amount of aid and other development assistance since the country had gone through massive destruction of infrastructure and other areas. During the '90s and 2000s,' the amount of aid and development assistance was reduced. However, the government has taken some large infrastructure projects in the past few years, and development assistance and aid has again risen to meet the country's development needs.

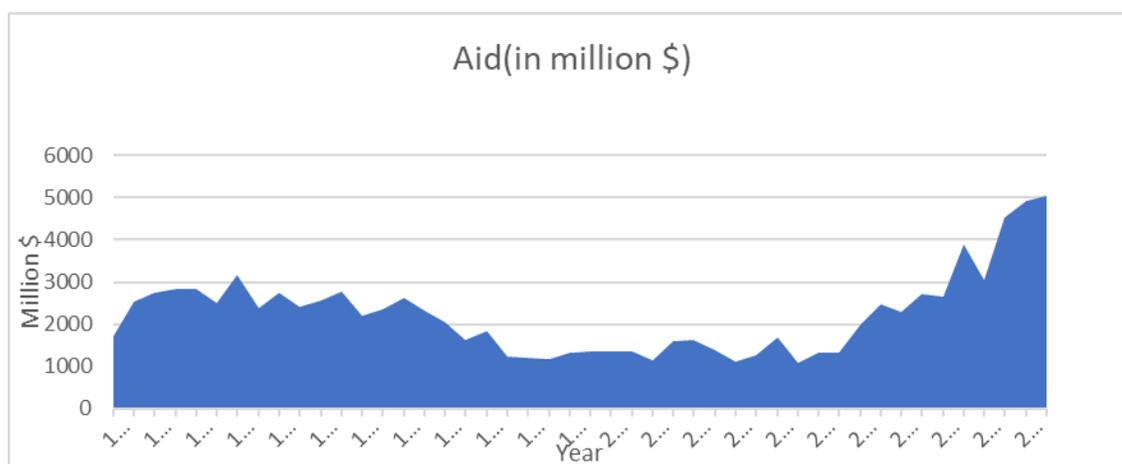

Figure-4: Aid Flows to Bangladesh Over 1976-2021

**Literature Review:**

Foreign Direct Investment (FDI) has been identified as one of the significant drivers of economic growth in many developing countries. FDI is the investment made by a foreign company in another country's economy. In recent years, an increasing interest has been in understanding the relationship between FDI and economic growth.

Many researchers have studied the impact of foreign capital on the economic growth of a country. The outcomes of these studies are mixed. Some found the impact positive, whereas some researchers found it negative. Some researchers also found a neutral relationship between foreign capital and economic growth. Most of the researchers focused on the impact of FDI on the economic growth of a country. Few researchers focused on remittance and aid in the economic growth of a country.

Researchers have identified that FDI may positively but indirectly impact economic growth when the host country can upgrade technology, invest in human capital, adapt trade policy and absorptive capacity. (Gönel and Aksoy, 2016; Katircioglu, 2009; Silajdzic and Mehic, 2016). FDI met the demand of capital formation in developing countries through capital investment, which ultimately enhance economic growth (Firebaugh, 1992). Mello (1999) is also found that FDI filled the resource shortfall in many developing countries. Fry (1999) identified that capital formation in South and East Asia is increased by FDI and thereby boosts economic growth.

Low-income countries grow at a greater rate due to the movement of technology and capital (Li and Chen, 2010). Foreign capital helps construct physical infrastructure like roads, factories and ports, which help the host country improve absorption capacity, attracting more FDI (Romer, 1993). Sridharan et al. (2009) showed that the economic development of developing countries is contributed by financial and economic adjustment together with stability of FDI.

Many researchers also found that countries with open trade policies benefit from the FDI inflows and the positive impact of FDI and trade is grasped when the economy is expected to grow faster and open economy (Adhikary, 2010;Shimul, Abdullah and Siddiqua, 2009). Open economy attracts multinational firms to invest in the host countries to acquire the benefits of the host countries like low-cost labor, natural resources etc. These multinational companies bring with them knowledge, managerial skills, ideas, and technologies to the host countries, which contribute to the economic development of the host countries (Mengistu and Adams, 2007).

Borensztein et al.(1998) conducted a study on FDI and economic growth, focusing on developing countries. They found a positive correlation between FDI and economic growth, suggesting that FDI can contribute to economic growth by providing additional capital, technology, and access to new markets.

Aitken and Harrison (1999) studied FDI on the productivity of domestic firms in developing countries. They found that FDI can have a positive spillover effect on the productivity of domestic firms, particularly in the manufacturing sector.

Blomstrom and Kokko (1998) studied the impact of FDI on economic growth in developing countries. They found that FDI can positively impact economic growth, particularly in countries with a high level of human capital and a stable macroeconomic environment.

Alfaro et al.(2004) examined the impact of FDI on economic growth in Latin American countries. They found that FDI can positively impact on economic growth, particularly in countries with a high level of financial development and a favorable investment climate.

Li et al. (2019) investigated the relationship between FDI and economic growth in the Belt and Road Initiative (BRI) countries. They found that FDI can positively impact on economic growth, particularly in countries with a high level of institutional quality and a favorable business environment.

Foreign Direct Investment (FDI) has been a topic of great interest for many years as it is believed to be an important driver of economic growth. However, there are some studies that suggest that FDI may have a negative impact on economic growth. Brecher and Alejandro (1977) showed that FDI has a negative impact on the host countries' economic growth if the multinational companies repatriate excess profit from the host country to the home country. Similarly, dependency theory also support the claim that FDI negatively affects the receiving country's economic growth (Dutt, 1997).

FDI crowding out domestic investment, which negatively affects the economic development of the receiving country. Eller et al. (2005) have examined 11 central and eastern European countries and found that FDI adversely affect economic growth by crowding out domestic investment.

Similarly, Blomstrom et al. (1996) examines the relationship between FDI and economic growth in a sample of 15 developing countries. The authors find that while FDI can have positive effects on economic growth, it can also have negative effects if it crowds out domestic investment.

Xu (2000) examines the relationship between FDI and economic growth in a sample of 40 countries. The author finds that FDI can have negative effects on economic growth if it leads to the displacement of domestic firms and the transfer of technology that is inappropriate for the host country's economic structure.

Aitken and Harrison (1999) examines the relationship between FDI and economic growth in a sample of Venezuelan manufacturing firms. The authors find that FDI negatively affects domestic firms' productivity, as it leads to increased competition and a diversion of resources away from domestic firms.

While FDI has the potential to promote economic growth, a growing body of literature suggests that it can also have negative impacts, such as crowding out domestic investment, creating a dual economy, and displacing domestic firms.

Remittances have been found to impact economic growth in many developing countries positively. Ratha and Mohapatra (2011) found that remittances have a more significant positive impact on economic growth in countries with low levels of financial development. This may be because in countries with limited financial development, remittances can help fill the gap in financing for investment and consumption needs. The influx of remittances can increase the economy's liquidity, leading to increased economic activity and growth.

On the other hand, Bazzi and Clemens (2013) found that remittances have a stronger positive impact on economic growth in countries with high levels of financial development. This may be because remittances in countries with well-developed financial systems can be more efficiently channeled into productive investments, such as entrepreneurship and small business development. In addition, remittances may help to increase access to credit and financial services, which can further stimulate economic growth.

Maimbo and Ratha (2005) found that remittance positively impact on economic growth in developing countries. The study showed that remittance increases recipient households' income, leading to increased consumption and investment. The increased consumption and investment lead to increased economic activity, which in turn leads to economic growth.

Aggarwal et al. (2011) found that remittance positively impact on financial development in developing countries. The study showed that remittance increases the amount of savings in recipient households, which leads to increased deposits in banks. The increased deposits in banks lead to increased financial intermediation and financial development. Several other literatures have also found positive and significant impact of remittance on the economic growth of a country (Ahmad et al, 2016; Azam, 2015; Comes et al, 2018; Cooray, 2012; Fayissa and Nsiah 2010; Hussain and Abbas, 2014; Jawaid and Syed Ali Raza, 2016; Meyer and Shera, 2017)

While remittance positively impacts economic growth in developing countries, there are also negative impacts associated with remittance. Remittance inflows can lead to the appreciation of the exchange rate, which can hurt the manufacturing sector and overall economic growth by making exports less competitive and imports cheaper, leading to a decline in the production of export goods. The study also highlights that remittances may encourage consumption instead of investment, and increase inequality and dependency (Chami, Fullenkamp and Jahjah, 2005).

Ratha (2003) found that remittances may have a negative impact on economic growth because they discourage recipients from participating in the labor force, leading to a decrease in the labor supply. This, in turn, may lead to a reduction in productivity and economic growth.

Karagoz (2009) identified negative correlation between remittance and economic growth in Turkey. Some researchers identified that remittance positively affects economic growth in the long run but in the short run there is a negative correlation between remittance and economic growth (Oshota and Badejo, 2014). However, Tolcha and Rao (2016) found that remittance has a negative and significant impact on economic growth in the short run but in the long run, the relation is positive and significant.

Pradhan (2016) studied the relationship between remittance and economic growth in Brazil, India, China, South Africa, and Russia. The result was mixed. Remittance has a positive impact on economic growth in China but in India, Brazil, and Russia, the relation is negative and significant. The relation is positive but insignificant in South Africa.

In addition to the above studies, Barajas et al. (2009) did not find any relation between remittance and economic growth in developing countries. Similarly, Shaikh et al. (2016) found that personal remittance has no impact on the economic growth of Pakistan.

Foreign aid refers to financial or technical assistance given by governments, international organizations, and individuals to developing countries. The debate over whether foreign aid promotes economic growth or not has been ongoing for decades. While some argue that aid has a negative impact on economic growth, others contend that aid has a positive impact on economic growth.

Yiew and Lau (2018) discovered that there is a U-shaped relationship between aid and economic growth. They found that aid negatively affected economic growth in the short run, but in the long run, aid positively affected economic growth in the selected countries. Similarly, Aghoutane and Karim (2017) identified that foreign aid encouraged economic growth in Vietnam in the short run but in the long run economic growth is negatively affected by foreign aid.

Some researchers also found that foreign aid is beneficial for a country's economic development to a particular level. Additional foreign aid adversely affects economic growth when the level is crossed (Fashina, 2018). Foreign aid stimulates economic growth to an optimum level because a large proportion of foreign aid is used for social welfare and humanitarian purpose (Pradhan and Phuyal, 2020).

However, some researchers have found that foreign aid has a negative impact on economic growth. Liaqat et al. ( 2019) empirically studied the impact of aid on the economic growth of some selected Asian countries over the period of 1984-2015. They found that aid negatively affects economic growth in these countries during the selected period. Similarly, Edward and Karamuriro (2020) identified that the relationship between foreign aid inflow and economic growth is negative for both short and long run.

Several previous literatures also investigate the relationship between FDI and economic growth, remittance, and economic growth and, Bangladesh's aid and economic growth. Reza et al., (2018) studied the impact of FDI on Bangladesh's economic growth from 1990-2015. They found that FDI positively relation with GDP for both the short and long run.

Faruk (2013) investigated the relationship between FDI and economic growth in Bangladesh and used data from 1980 to 2011. He also found a positive relation between FDI and economic growth. Similarly, Noor et al. (2016) and Faruq (2023) also found a positive relationship between FDI and economic growth.

Ahamad and Tanin (2010) identified that GDP growth is an important determinant of FDI. They found that foreign investors opted to invest in large markets and a country with a high economic growth rate. However, Shimul et al. (2009) found that GDP per capita was caused by FDI and FDI and GDP was not cointegrated. Tabassum and Ahmed (2014) also found that FDI is insignificant in affecting economic growth of Bangladesh. Similarly, Faisal and Islam (2022) found negative relation between FDI and GDP growth in the short run and no causality between FDI and GDP.

Siddique et al. (2012)investigated the impact of remittance on the economic growth of Bangladesh, India, and Sri Lanka over 25 years. They found that remittance has a positive impact on the economic growth of Bangladesh. Azam (2015) also investigated the impact of workers' remittance on the economic growth of four Asian developing countries from 1976-2012. He found a significant positive relationship between workers' remittance and economic growth.

Sutradhar (2020), however, found that remittance has a negative effect on the economic growth in Bangladesh when She investigated the impact of remittance on the economic growth in Bangladesh, India, Pakistan, and Sri Lanka over the period of 1977-2016.

Islam(1992) investigated the impact of foreign aid on the economic growth of Bangladesh and found that aid had no significant contribution on economic growth. However, Golder et al. (2021) studied the impact of foreign aid on the economic growth of Bangladesh over the period from 1989 to 2018. They found a robust and significant impact of foreign aid on the economic growth of Bangladesh.

## Methodology:

*Data and Model*

This study uses time series data (1976-2021) on four variables for Bangladesh- Gross Domestic Product (GDP), Foreign Direct Investment (FDI), Remittance and Official Development Assistance and Aid. All the data were collected from the World Development Indicator (WDI) of the World Bank. The time series data is taken for 46 years including most of Bangladesh's economic period since Bangladesh got independence in 1971. Most of the previous studies include only FDI or remittance or aid to measure their impact on the economic growth in Bangladesh. The results are also mixed. Some found a positive impact, whereas others found a negative impact of the same variable on the economic growth in Bangladesh. Therefore, this study includes all three sources of foreign capital inflows in Bangladesh to measure their respective impact on the economic growth. Moreover, this study also creates an index named foreign capital index by principal component analysis (PCA) to measure the impact of foreign capital on the economic growth in Bangladesh.

Table- 1: Description of the Variables:

Variables    Descriptions

| Variables | Descriptions |
|---|---|
| lnGDP | Log of GDP used as a proxy to economic growth |
| lnFDI | Log of FDI (net inflows) |
| lnrem | Log of Personal Remittance received |
| lnaid | Log of Net Official Development Assistance and Official Aid received |
| FCDI | Foreign Capital Depthness Index |
| μ | Error term |

Several models have been developed (Golder et al., 2021) to examine the impact of FDI, remittance and aid on the economic growth of Bangladesh. Details are given in table 1. Foreign capital depthness index is created by Principal Component Analysis (PCA) using the three components of foreign capital: FDI, remittance and aid. This index is used to ascertain the joint impact of these three foreign capital components on Bangladesh's economic growth.

$$lnGDP = \beta_0 + \beta_1 lnFDI + \beta_2 lnrem + \beta_3 lnaid + \mu \qquad (1)$$

*Unit Root Tests*

Augmented Dickey-Fuller (ADF) and Phillips-Perron (PP) unit root tests are used to examine whether they are stationary at level. If the variables are non-stationary at level, the difference form of the variables is used to examine whether they are stationary at difference form. The null hypothesis is the time series has a unit root which means that the series is not stationary. The alternative hypothesis is that the time series has no unit root which means that the series is stationary. Both ADF and PP tests examined the presence of unit root in the time series. The null hypothesis has to be rejected at level or difference to be stationary. The variables may be stationary at level, that is, I (0) or they may be stationary at first difference, that is, I(1).

Table- 2: ADF Unit root Tests

| Variables | Levels test statistics | 1ST Difference Test Statistics | Order of Integration |
|---|---|---|---|
| lnGDP | 0.365 | -4.350 *** | I (1) |
| FCDI | -0.015 | -4.150 *** | I (1) |
| lnFDI | -2.494 | -6.725 *** | I (1) |
| lnAid | -1.521 | -4.286*** | I (1) |
| lnRem | -1.485 | -4.231*** | I (1) |

The lagged value of the dependent variable is added to the Dickey-Fuller(DF) equation and thereby carried out the ADF test (Dickey and Fuller, 1979). The Dickey-Fuller equation is:

$$\Delta Y_t = \beta_1 + \beta_2 t + \rho Y_{t-1} + \mu_t \qquad (2)$$

The ADF equation is:

$$\Delta Y_t = \beta_1 + \beta_2 t + \rho Y_{t-1} + \alpha_I + \sum_{i=1}^{m} \Delta Y_{t-1} + \varepsilon_t \qquad (3)$$

Equation 2 of DF is augmented to equation 3 by adding the dependent variable's lag value, which is, $\Delta Y\_t= (Y\_(t-1)-Y\_(t-2))$. $\varepsilon\_t$ is the error term. The presence of the unit root in the time series is tested in the ADF like DF.

The PP test uses nonparametric correlation (Newey and West, 1994) for possible serial correlation of the error term instead of including lagged difference terms of the dependent variables. The simple OLS used by Phillips and Perron (1988) is:

$$\Delta Y_t = \mu + \alpha Y_{t-1} + \mu_t \qquad (4)$$

This model tests the presence of the unit root in the time series. The null hypothesis is the presence of unit root in the time series and the alternative hypothesis is no unit root in the time series, that is, whether the data is stationary or non-stationary.

The test results in Table 2 show that all the variables are non-stationary at level and stationary at first differences in ADF test. However, PP test in Table 3 shows that lnFDI and lnRem are stationary in both level and first difference.

Table- 3: Phillips-Perron Unit root Tests

| Variables | Levels test | 1ST Difference Test | Order of Integration |
|---|---|---|---|
| lnGDP | 0.896 | -5.841*** | 1(1) |
| FCDI | -0.182 | -7.142*** | 1(1) |
| lnFDI | -2.905** | -9.769 *** | 1(1) |
| lnAid | -2.158 | -9.729*** | 1(1) |
| lnRem | -3.377** | -8.819*** | 1(1) |

Lag Length Selection Test and Results

The Johansen cointegration test investigates the long-run relationship between the dependent and independent variables. Before the cointegration test, proper lag length selection is required. Different lag length criteria are used to determine the optimal lag length: final prediction error (FPE), Akaike's information criterion (AIC), Hannan and Quinn information criterion (HQIC), and Schwarz's Bayesian information criterion (SBIC) (Akaike, 1974). The following equation is used to determine FPE.

$$FPE = |\Sigma_u|((T + \bar{n}|T - \bar{n}))^k \quad (5)$$

Similarly, the computation of other information criteria is based on their respective standard definition.

In this equation, T refers to number of observations and $\bar{n}$ refers to the average number of parameters over k equations.

Table- 4: Lag Length Selection

| lags | LL | LR | FPE | AIC | HQIC | SBIC |
|---|---|---|---|---|---|---|
| 0 | -191.35 | NA | .128855 | 9.3024 | 9.36306 | 9.4679 |
| 1 | -2.67844 | 377.34 | .000035 | 1.07993 | 1.38322* | 1.90739* |
| 2 | 17.1955 | 39.748 | .00003* | .895454* | 1.44139 | 2.38489 |
| 3 | 32.8438 | 31.297* | .000032 | .912201 | 1.70077 | 3.0636 |
| 4 | 45.661 | 25.635 | .000041 | 1.06376 | 2.09497 | 3.87713 |

* lag order selected by particular criterion.

LL= log likelihood; LR= log ratio; FPE= final prediction error; AIC= Akaike's information criterion; HQIC= Hannan and Quinn information criterion; SBIC= Schwarz's Bayesian information criterion.

Table 3 shows different lag length criteria supporting different optimum lag length. In this study we use a lag length of 2 as suggested by final prediction error (FPE) and Akaike's information criterion (AIC). According to Johansen and Juselius (1990) for a small sample size, maximum lag length should be two.

*Cointegration Test:*

The Johansen cointegration test is carried out to ascertain the long-run relation between dependent and independent variables. This test presents the cointegration vectors. Johansen and Juselius (1990) recommend two statistic tests for determining the number of cointegrating vectors. These two tests are: trace test and the maximum eigenvalue test.

The null hypothesis of trace test is that the time series has r cointegrating vectors where the alternative hypothesis is that there is r+1 vectors. If the critical value is less than the test statistic, then the alternative hypothesis will be accepted. The trace test statistic equation is:

$$\lambda_{trace} = -T \sum_{i=r+1}^{d} \ln(1 - \lambda_i) \qquad (6)$$

Similarly, r number of cointegrating vectors is the null hypothesis of the maximum eigenvalue test where more than r number of cointegrating vectors is the alternative hypothesis. The equation of the maximum eigenvalue is:

$$\lambda_{max} = -T \ln(1 - \lambda_{r+1}) \qquad (7)$$

Similar to the trace test, the null hypothesis is rejected if the test statistic is greater than the critical value given in the Johansen table. The result in Table 5 shows that there exists at least one cointegrating vector between the variables.

Table- 5: Johansen Cointegration Test Results

| Hypothesized no. of Cointegrating Equations (CEs) | Trace | | Maximum eigenvalue | |
|---|---|---|---|---|
| | Trace statistic | Critical value (5%) | Max-eigen statistic | Critical value (5%) |
| None | 54.3045 | 47.21 | 41.7304 | 27.07 |
| At most 1 | 12.5741* | 29.68 | 8.6406 | 20.97 |
| At most 2 | 3.9335 | 15.41 | 3.1966 | 14.07 |
| At most 3 | 0.7369 | 3.76 | 0.7369 | 3.76 |

* Indicates rejection of the null hypothesis at 5% significance level that is no cointegration between the variables.

### Vector Error Correction Model (VECM)

The Johansen cointegration test identified in this study that the variables have a long-term association since both the trace test and the maximum eigenvalue test found at least one cointegration among the variables. Since the variables are cointegrated, we use VECM instead of VAR model since VAR model is used when variables are not cointegrated.

Table- 6: Johansen normalization restriction imposed (Long Run Causality)

| beta | Coef. | SE | z | P>|z| | [95% Conf. | Interval] |
|---|---|---|---|---|---|---|
| ce1 | | | | | | |
| lngdp | 1 | . | . | . | . | . |
| lnfdi | -.1609077 | .0190216 | -8.46 | 0.000 | -.1981893 | -.123626 |
| lnrem | -.0868698 | .0668819 | -1.30 | 0.194 | -.2179559 | .0442162 |
| lnaid | -.6901144 | .1501995 | -4.59 | 0.000 | -.9845 | -.3957287 |
| cons | -5.40912 | . | . | . | . | . |

Table 6 represents the error term. The normalization is on lngdp since it is the outcome variable. The signs of the coefficient will be reversed in the long run by taking the vector to be zero as a cointegrating vector can reformulate the long run equation. The findings of Table 6 show that, In the long run, lnfdi and lnaid have positive and significant Impact on lngdp. The coefficients are statistically significant at the 1% level. However, lnrem has a positive but insignificant impact on lngdp. Therefore, lnfdi and lnaid have asymmetric effects on lngdp in the long run, on average, ceteris paribus. Therefore, the error correction term can be written as:

$$ECT_{t-1} = 1.00 lngdp_{t-1} - 0.161 lnfdi_{t-1} - 0.0869 lnrem_{t-1} - 0.690 lnaid_{t-1} - 5.401$$

Table-7: Johansen normalization restriction imposed (for fcdi)

| beta | Coef. | SE | z | P>|z| | [95% Conf. | Interval] |
|---|---|---|---|---|---|---|
| _ce1 | | | | | | |
| lngdp | 1 | . | . | . | . | . |
| fcdi | -2.211841 | .7866318 | -2.81 | 0.005 | -3.753611 | -.6700708 |
| _cons | -5.40912 | . | . | . | . | . |

Table 7 represents the creation of the error term of the foreign capital depthness index. In this study, foreign capital depthness index (FCDI) is created by using data on FDI, remittance and aid through principal component analysis (PCA). Table 7 shows that foreign capital depthness index (FCDI) has a positive and significant impact on economic growth (lngdp) in the long run.

*OLS Estimation:*

Ordinary Least Square (OLS) method can provide valid results since the cointegration test results confirm that at least one cointegrating vector exists between the variables (Rana and Wahid, 2017).

Table- 8: Results of Ordinary Least Squares Test

| Variables | Coef. | SE | t-value | p-value |
|---|---|---|---|---|
| lnfdi | .025 | .01 | 2.54 | .015 |
| lnrem | .519 | .031 | 16.78 | 0 |
| lnaid | .393 | .083 | 4.72 | 0 |
| Constant | 4.761 | 1.774 | 2.68 | .01 |

*** $p<.01$, ** $p<.05$, * $p<.1$
Lngdp is the dependent variable. Value of R-square and adjusted R-square are 0.9412 and 0.9370, respectively. The value of F statistic is 224.00, and the p value of F statistic is 0.

Table 8 shows the results of OLS. From the results we can write the econometric model of equation 1 as follows:

lngdp= 4.761+ 0.25×lnfdi + .519×lnrem + .393×lnaid + μ          (8)

The results of OLS indicate that 1% increase in FDI increases 0.025% GDP. Similarly, 1% increase in remittance increases GDP by 0.519% and 1% increase in aid increases GDP by 0.393%.

Table- 9. Results of Ordinary Least Squares Test (for fcdi)

| Variables | Coef.  | SE   | t-value | p-value |
|-----------|--------|------|---------|---------|
| fcdi      | 2.816  | .258 | 10.93   | 0       |
| Constant  | 23.946 | .107 | 224.83  | 0       |

*** $p<.01$, ** $p<.05$, * $p<.1$

Table 9 shows that the foreign capital depthness index significantly impacts Bangladesh's economic growth. A 1% increase in foreign capital can lead to 2.816% increase in GDP. The above analysis indicates that the individual impact of FDI, remittance, and aid is lower than the combine impact of these variables on the economic growth of Bangladesh.

Table- 10: Results of Coefficient Diagnostics (Wald Statistics)

Dependent variable: D(LNGDP)

| Excluded | Chi-sq | df | Prob. |
|---|---|---|---|
| D(LNFDI) | 3.866195 | 2 | 0.1447 |
| D(LNREMITTANCE) | 0.216482 | 2 | 0.8974 |
| D(LNAID) | 0.360777 | 2 | 0.8349 |
| All | 4.162579 | 6 | 0.6547 |

Dependent variable: D(LNFDI)

| Excluded | Chi-sq | df | Prob. |
|---|---|---|---|
| D(LNGDP) | 28.19226 | 2 | 0.0000 |
| D(LNREMITTANCE) | 35.17650 | 2 | 0.0000 |
| D(LNAID) | 3.047518 | 2 | 0.2179 |
| All | 80.04162 | 6 | 0.0000 |

Dependent variable: D(LNREMITTANCE)

| Excluded | Chi-sq | df | Prob. |
|---|---|---|---|
| D(LNGDP) | 8.948086 | 2 | 0.0114 |
| D(LNFDI) | 2.124926 | 2 | 0.3456 |
| D(LNAID) | 1.131313 | 2 | 0.5680 |
| All | 13.60696 | 6 | 0.0343 |

Dependent variable: D(LNAID)

| Excluded | Chi-sq | df | Prob. |
|---|---|---|---|
| D(LNGDP) | 0.878907 | 2 | 0.6444 |
| D(LNFDI) | 0.729678 | 2 | 0.6943 |
| D(LNREMITTANCE) | 0.528325 | 2 | 0.7678 |
| All | 2.715631 | 6 | 0.8436 |

We have used the Wald test here to justify the VECM model. Table 10 represents the results of the Wald test. Table 10 indicates that there is no short-run causality between FDI and economic growth, remittance and economic growth, and aid and economic growth. However, economic growth can affect FDI and remittance in the short run but it has no relation with aid in the short run. Moreover, there is a presence of unidirectional causality from remittance to FDI; that is, remittance causes FDI but FDI does not cause remittance in the short run. Aid has no causal relation with FDI, remittance and GDP in the short run. When we use the foreign capital depthness index (fcdi) to measure its relationship with economic growth in the short run we find that GDP has no causal relationship with fcdi and fcdi has no causal relationship with GDP in the short run.

Diagnostic Tests:

We employ some diagnostic tests to ensure that the model is well specified and assumptions are valid. These tests are conducted to check whether the residuals from the VECM satisfy certain properties that are required for the model to be valid. These tests are presented in table 11 and table 12.

Table- 11: Autocorrelation Test

| lag | chi2 | df | Prob>Chi2 |
|---|---|---|---|
| 1 | 12.629 | 16 | 0.742 |
| 2 | 11.864 | 16 | 0.753 |

Table- 12: Normality Test

| Equation | chi2 | df | Prob > chi2 |
|---|---|---|---|
| D_lngdp | 6.652103 | 2 | 0.3595 |
| D_lnfdi | 1.698420 | 2 | 0.4278 |
| D_lnrem | 1.004480 | 2 | 0.6052 |
| D_lnaid | 1.240806 | 2 | 0.5377 |
| ALL | 10.59581 | 8 | 0.2257 |

This study uses the Lagrange Multiplier (LM) test to identify the autocorrelation of the residuals. The goodness of fit in the time series analysis is dependent on the no serial correlation in the residuals. In this study we have found that LM test can't reject the null hypothesis of no serial correlation among the residuals. Therefore, the residuals are not correlated. In addition, we use the Jarque-Bera test to identify whether the errors are normally distributed. The Jarque-Bera test shows that overall, the errors are normally distributed as well as individually.

## Conclusions

This study attempts to identify the impact of foreign capital on the economic growth of Bangladesh using time series analysis. Foreign capital includes FDI, remittance, and aid. Using Principal Component Analysis (PCA), this study develops an index named foreign capital depthness index based on data collected from World Development Indicators (WDI) on FDI, remittance, and aid. This study is consistent with many other previous studies on the economic growth of Bangladesh. However, this study is different from previous studies where individually showed the impact of FDI, remittance and aid on the economic growth of Bangladesh but this study builds an index and shows the combined impact of FDI, remittance and aid on the impact of Bangladesh.

The ADF and PP tests are used to identify the stationarity of the variables and found that all the variables are stationary at first difference. The Johansen cointegration test is used since all the variables are stationary at first difference. The optimum lag length is selected before running the Johansen cointegration test as suggested by AIC and FPE. The Johansen cointegration test identified least 1 cointegration exists among the variables. Vector Error Correction Model (VECM) is used since cointegration exists among the variables.

The results from VECM show that FDI, remittance, and aid positively impact Bangladesh's economic growth in the long-run. However, they have no short-run causality to economic growth. Similarly, foreign capital depthness index (fcdi) has a positive impact on the economic growth of Bangladesh in the long run but it has no relation with GDP in the short run. The results, however, show that fcdi has a greater impact on GDP than the individual impact of FDI, remittance, and aid on GDP. FDI is significant for the economic growth of many developing countries like Bangladesh. However, remittance, official development assistance, and aid have negatively impacted the economic growth of many developing countries. This study's findings show that the combined impact of all the variables of foreign capital is higher on the economic growth in Bangladesh. Therefore, the government of Bangladesh should ensure the smooth flow of all these three variables for the higher economic growth of Bangladesh rather than focusing on the individual components of the foreign capital.